\documentclass[aps,prb,twocolumn,showpacs,letterpaper, superscriptaddress]{revtex4}
\usepackage{graphicx}
\begin{document}

\title{Magnetism, structure, and charge correlation at a pressure-induced Mott-Hubbard insulator-metal transition}
\author{Yejun Feng}
\affiliation{The Advanced Photon Source, Argonne National Laboratory, Argonne, IL 60439, 
USA}
\affiliation{The James Franck Institute and Department of Physics, The University of Chicago, Chicago, IL 60637, USA}
\author{R. Jaramillo} 
\affiliation{School of Engineering and Applied Sciences, Harvard University, Cambridge, MA 01238, USA}
\author{A. Banerjee}
\affiliation{The James Franck Institute and Department of Physics, The University of Chicago, Chicago, IL 60637, USA}
\author{J. M. Honig}
\affiliation{Department of Chemistry, Purdue University, West Lafayette, IN 47907, USA}
\author{T. F. Rosenbaum} 
\affiliation{The James Franck Institute and Department of Physics, The University of Chicago, Chicago, IL 60637, USA}

\begin{abstract}
We use synchrotron x-ray diffraction and electrical transport under pressure to probe both the magnetism and the structure of single crystal NiS$_{2}$ across its Mott-Hubbard transition. In the insulator, the low-temperature antiferromagnetic order results from superexchange among correlated electrons and couples to a (1/2, 1/2, 1/2) superlattice distortion. Applying pressure suppresses the insulating state, but enhances the magnetism as the superexchange increases with decreasing lattice constant. By comparing our results under pressure to previous studies of doped crystals we show that this dependence of the magnetism on the lattice constant is consistent for both band broadening and band filling. In the high pressure metallic phase the lattice symmetry is reduced from cubic to monoclinic, pointing to the primary influence of charge correlations at the transition. There exists a wide regime of phase separation that may be a general characteristic of correlated quantum matter.
\end{abstract}

\pacs{71.30.+h, 75.30.Kz, 61.05.cp,  64.75.Qr }


\maketitle

\section {Introduction}
One of the great challenges of understanding correlated materials is teasing apart the relative influences of the spin, charge, orbital, and lattice degrees of freedom. The point where an insulator becomes a metal highlights acutely the competition between mechanisms, but it also affords a special opportunity to limn the pertinent physics when different routes across the phase transition are available. Going back to the original ideas of Mott and Hubbard, we know that strong Coulomb repulsion between electrons on a single lattice site can localize charge even when band theory predicts metallic behavior \cite{2, 6}, and antiferromagnetism was attributed to a consequence of superexchange between localized electrons \cite{24}. At the same time, Slater claimed that antiferromagnetism alone could account for the formation of the insulating gap \cite{3}. All can be subsumed by symmetry changes wrought by a structural phase transition \cite{5}. 

The cubic pyrite crystal NiS$_2$ has long been recognized as a canonical Mott-Hubbard correlated insulator \cite{6, 7, 9, 10, 13, 14, 15}. Band structure calculations \cite{14} put the sulfur $3p$ band and Ni $t_{2g}$ band well below the half filled Ni $e_g$ band, pointing to on-site Coulomb repulsion as the source of the insulating energy gap $E_g$, which lies in the range 1-10 meV (Ref. [\onlinecite{15}]). The small size of this gap demonstrates that NiS$_2$ is an incipient Mott insulator \cite{13} with the Coulomb repulsion $U$ comparable to the $e_g$ bandwidth, $W$ = 2.1 eV (Ref. [\onlinecite{14}]). The gap can be suppressed either by Se doping \cite{7, 11, 12} or applied pressure \cite{16, 17, 19}, but the scale of the pressure required to drive the gap to zero in the pure limit has introduced technical obstacles to systematic studies of the competition between electronic, magnetic, and structural correlations at the quantum phase transition. Doping with Se expands the lattice and reduces the Ni $3d$ bandwidth \cite{2, 7}, and the transition in this case is thought to be driven by increasing charge transfer between the Ni $3d$ and Se $4p$ bands \cite{6}. Applying pressure tunes the ratio $U/W$ and provides a more direct approach to the Mott-Hubbard model. 

We use synchrotron x-ray diffraction and electrical transport in a diamond anvil cell to parse the roles of the low temperature antiferromagnetism and the lattice structure, both through the insulating state and at the transition to the metal in the Mott-Hubbard system, NiS$_{2}$. The reduced symmetry in the metal to monoclinic - a highly unusual occurrence for correlated materials described below - eliminates the change in the structure as a likely origin of the delocalization of charge.  By comparison of the pressure-induced transition in the pure compound to the insulator-metal transition driven by chemical substitution of Se for S, we identify the charge degrees of freedom as the predominant driving mechanism. Realizing the insulator-metal transition in high quality single crystals of a stoichiometric material using applied pressure further clarifies the physics by avoiding complications that arise from chemical disorder, most notably the competition between Anderson localization and the Mott transition. 

\section {Experimental Methods}
High-pressure single-crystal x-ray diffraction measurements were carried out at beamlines 4-ID-D and 6-ID-B of the Advanced Photon Source. In a vertical scattering geometry with  a psi-diffractometer, a high \textit{q}-resolution (FWHM  $\sim1\times 10^{-3} \mathrm{\AA}^{-1}$) is achieved using a 50 $\mu m$ size detector slits positioned 1.3 m away from the sample along the $2\theta$ arm \cite{29}. The use of double-bounce Pd mirrors for 20 keV x-rays and an energy discriminating NaI scintillation detector eliminated higher-harmonic contamination of the diffraction signal. Our single crystals were grown by the Te flux method to remove potential complications from excess impurity concentrations \cite{11}. Crystals were 25 to 50  $\mu m$ in diameter and fit well within the diamond anvil cell pressure chamber. Five different crystals were studied under pressure using a methanol:ethanol 4:1 mixture for the pressure medium. Base temperature varied between 3.5 and 5.8 K, and the pressure was calibrated \textit{in situ} using silver diffraction \cite{29}. 

The problem of vacancies, common to sulfides, is well characterized  \cite{10} in NiS$_2$. Both S (about 4\%) and Ni (varying) vacancies can be determined from the measured lattice constants and electrical resistivity; in this way we estimate our sample stoichiometry to be NiS$_{1.96}$. There are two coexisting antiferromagnetic structures in NiS$_2$. The M1 antiferromagnetic order with a wavevector (1, 0, 0) emerges from a second order phase transition at $T_{N1}$ = 37 to 54 K, where $T_{N1}$ strongly depends on vacancy concentration and varies from sample to sample (\textit{i.e.} Ref. [\onlinecite{9}]). The M2 antiferromagnetic order has a (1/2, 1/2, 1/2) wavevector and emerges at a first order transition at $T_{N2} = 30 \pm 1$ K, where the transition temperature is consistent across all published reports including those for different vacancy concentrations. However, vacancies are responsible for the variable canting angle in an antiferromagnet \cite{30}, as observed in the M2 phase of NiS$_2$ (Ref. [\onlinecite{9, 10}]). 

\section {Antiferromagnetism at ambient and high pressures}
One outstanding question in the field of Mott-Hubbard systems concerns the role of period doubling antiferromagnetism at the insulator-metal transition.  The magnetostrictive response of the lattice to M2 is likely rhombohedral \cite{20} and the distortion from cubic symmetry is extremely small, with a relative lattice constant change $\Delta a/a \sim 2 \times 10^{-4}$ (Ref. [\onlinecite{21}]). Therefore, for our diffraction measurements, we were able to model the insulating ground state using a cubic matrix. At ambient pressure and $T < T_{N2}$ we report the discovery of charge-originated superlattice diffraction peaks at (1/2, 1/2, 1/2)-type positions in reciprocal space (Fig. 1), corresponding to the M2 magnetic structure. Given the compatibility of diamond anvil cell technology with high energy x-ray diffraction, this opens up the possibility of combined magnetic and structural studies of the pressure-driven Mott-Hubbard transition in pure NiS$_2$. The temperature dependence of the superlattice intensity (Fig. 1a) scales linearly with the magnetic M2 neutron diffraction intensity \cite{9}. The x-ray supperlattice and neutron M2 diffraction intensities scale quadratically with the superlattice displacement and also the magnetic moment, respectively.  The scaling in Fig. 1a thus points to a linear coupling between the M2 magnetic moment and the superlattice displacement. A linear coupling between magnetism and the lattice is rather common in solids, and most often observed in low-dimensional systems \cite{22}. By comparison, the external magnetostriction \cite{21, 32}, measured by the linear thermal expansion, scales quadratically with the M2 magnetic moment (Fig. 1a). 

\begin{figure}
\begin{center}
\includegraphics[width=3.3in]{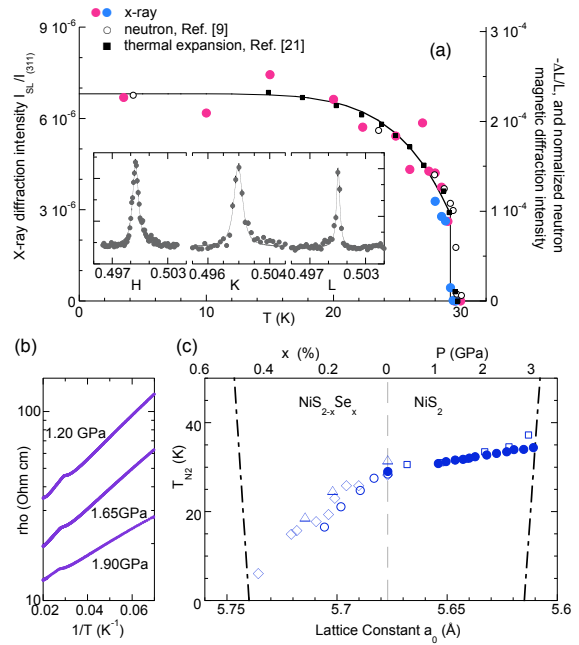}
\caption{(color online). (a) Superlattice intensity $I_{SL}$ at ambient $P$ plotted as a function of temperature. $I_{SL}$ is averaged over (3.5, 1.5, 1.5), (3.5, 0.5, 1.5), (2.5, 1.5, 1.5), and (2.5, 0.5, 1.5), and normalized to (3, 1, 1). Data were taken while warming (pink) and cooling (blue) through $T_{N2}$ = 29.2 K. The x-ray superlattice intensity scales linearly with the intensity of magnetic neutron diffraction from the (1/2, 1/2, 1/2) M2 magnetic order (open circles, data from Ref. [\onlinecite{9}]), as well as the linear thermal expansion $|\Delta L/L|$ along (1, 1, 1) (black solid squares, data from Ref. [\onlinecite{21}]). Solid line is a guide to the eye. (inset) H, K, L scans of (1/2, 1/2, 1/2) superlattice diffraction at ambient $P$ and $T$ = 5.8 K. (b) Resistivity of NiS$_2$ at several pressures. All data follow an Arrhenius form through the first order phase transition at $T_{N2}$, which is marked by a kink. (c) \textit{P-x-T} phase diagram of the M2 magnetic phase. $T_{N2}$ is plotted as a function of low temperature cubic lattice constant $a_0$ for both NiS$_2$ under pressure and NiS$_{2-x}$Se$_{x}$ (blue open symbols: [\onlinecite{12, 16, 19}]; blue solid circle: our transport measurements). The two dot-dash lines mark the first-order boundaries of the insulator-metal transition driven by pressure and chemical substitution, and bound the M2 order. }
\end{center}
\end{figure}

\begin{figure}
\begin{center}
\includegraphics[width=3.2in]{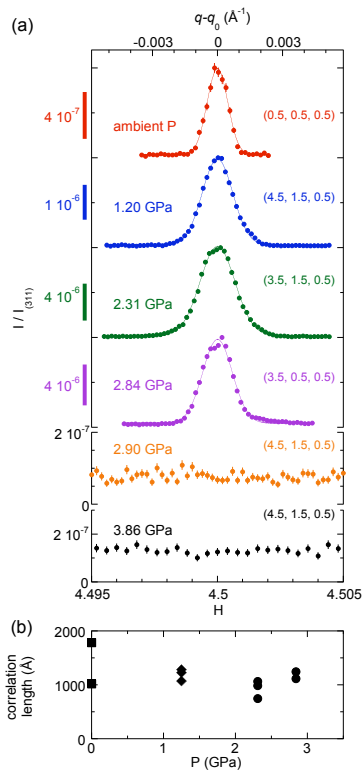}
\caption{(color online).  (a) Comparison of the high-resolution longitudinal linescans of the superlattice at four pressures between 0 and 2.84 GPa. Data are plotted from the center $q_0$ of each individual peak and displaced vertically with scale bars representing individual intensity relative to that of the (3, 1, 1) order. All lineshapes are nearly resolution limited approaching the phase boundary. This indicates that the crystal remains cubic with long M2 correlation lengths throughout the insulating phase. The antiferromagnetic M2 order disappears at higher pressures as exemplified by two null scans through the superlattice positions at $2.90 \pm 0.05$ and $3.86 \pm 0.12$ GPa. (b) Calculated lower bounds of correlation lengths for the superlattice. Separate points at a given $P$ represent scans through different superlattice orders. All measurement temperatures were between 3.5 and 5.8K. }
\end{center}
\end{figure}

The intensity of the superlattice peaks provides a quantitative estimate of lattice distortion {\boldmath$\delta$} through the relation $I_{SL}/I_{(311)}=|\mathbf{q}\cdot${\boldmath$\delta$}$/2|^{2}S(SL)/S(311)$, where $\mathbf{q}$ is the superlattice ordering wavevector and $S(SL)$ and $S(311)$ are the atomic form factors at the superlattice and (3, 1, 1) positions, respectively. Given the domain degeneracy in each superlattice order and uncertainty over the direction of {\boldmath$\delta$}, we evaluate $|\mathbf{q}\cdot${\boldmath$\delta$}$/2|^{2}$  by averaging the direction of superlattice displacement {\boldmath$\delta$} uniformly over the full $4\pi$ solid angle. Further averaging was carried out over the four (1/2, 1/2, 1/2)-type superlattice domains. In this way we estimate $\delta = 2.3 \pm 0.7~10^{-3} \mathrm{\AA}$, which is a factor of  $4\times10^{-4}$ smaller than the lattice constant $a$. Using a value for the lattice force constant appropriate to pyrite-structured $3d$ transition metal dichalcogenides ($\sim1$ N/cm = 6.25 eV/ $\mathrm{\AA}^2$, Ref. [\onlinecite{23}]), we estimate that the elastic energy associated with the superlattice distortion is approximately 0.017 meV per Ni atom. This is at least two orders of magnitude smaller than the insulating gap and the magnetic exchange coupling, and we therefore consider the superlattice to be a minimally intrusive representation of the underlying magnetic order. We also point out that the superlattice distortion cannot account for the insulating behavior. For a wide band model appropriate to NiS$_2$ ($W \gg E_{g}$) both the lattice and electronic energies scale as $\delta^2$, and we need only confirm that net quadratic coefficient is positive. Using Eq. (3.54) of Ref. [\onlinecite{5}] and inserting appropriate values we find that the quadratic coefficient lies in the range 2.4 - 3.1 eV/$\mathrm{\AA}^2$. Rather than coupling to the formation of an energy gap, the superlattice is likely driven by the variation in exchange constant $J$ with Ni ion displacement \cite{22}.

Using the superlattice reflections as a measure of the M2 order, we are able to track the magnetism and the crystal lattice through the pressure-driven insulator-metal transition. The only high-pressure magnetic scattering study published to date by Panissod \textit{et al.} (Ref. [\onlinecite{16}]) was limited to $P < 2.9$ GPa at 4.2 K with no disappearance of antiferromagnetism observed. This study identified the insulator-metal phase boundary with a suspected lattice discontinuity at $1.3 \pm 0.4$ GPa, leading to the conclusion that the magnetism is continuous across the insulator-metal transition \cite{16}. Here we observe the (1/2, 1/2, 1/2)-type superlattice distortion at every pressure from 0 to 2.84 GPa (Fig. 2a) at base temperature. Above 2.84 GPa the superlattice vanishes in all samples (Fig. 2c-d). Published accounts \cite{17, 19} and our own transport measurements place the critical pressure for the insulator-metal phase boundary in the range 2.2 - 3.1 GPa. This range brackets the upper limit of the observed superlattice diffraction. It is therefore natural to conclude that the disappearance of the low temperature M2 magnetic state, the structural phase transition (see below), and the insulator-metal transition all coincide at $P \sim 2.9$ GPa.

The relationship between the correlated insulator and M2 antiferromagnetism is further revealed by considering the phase boundary $T_{N2}$ as a function of lattice constant for both NiS$_2$ under pressure and NiS$_{2-x}$Se$_{x}$ in the \textit{P-x-T} phase diagram (Fig. 1c). Antiferromagnetic coupling of correlated electrons results from superexchange through the S ligand fields \cite{24}. This coupling grows stronger as the lattice constant is reduced \cite{2}, whether by chemical substitution or by applied pressure, continuing smoothly and continuously across the substitution-pressure interface. For the NiS$_{2-x}$Se$_{x}$($P$, $x$) system, $T_{N2}$ increases as the lattice constant shrinks, consistent with the superexchange interaction. The M2 magnetism and the associated superlattice distortion should be considered as byproducts of electron correlation and are not by themselves responsible for driving the insulating state. Important evidence for this also comes from electrical resistivity data which show that the Arrhenius activation energy is unchanged on cooling through $T_{N2}$ (Fig. 1b). The insulating energy gap is thus well established before the formation of the M2 phase, reflecting the fact that the energy scales of the charge and magnetic interactions are well separated. The Hubbard $U$ is comparable to the Ni $3d$ $e_g$ bandwidth $W= 2.1$ eV, the magnetic exchange coupling is comparable to the transition temperature $k_{B}T_{N2} = 2.6$ meV, and the superlattice distortion energy scale is an even smaller 0.02 meV.

Careful study of the lattice structure reveals additional information about the nature of the transition. High-resolution $\theta$ - $2\theta$ longitudinal scans at all orders of the superlattice reflection are consistently close to resolution-limited (Fig. 2a), giving M2 correlation lengths greater than 1000 \AA (Fig. 2b). This is consistent for all pressures within the phase boundary. For the fcc lattice peaks, longitudinal scans (Fig. 3) reveal a more complicated picture. All measured lineshapes are resolution-limited for $0 < P < 0.85$ GPa. However, beginning at 0.85 GPa we observe multiple splitting of diffraction peaks in almost every sample at every pressure. This indicates the emergence of structural domains of reduced symmetry. Importantly, the superlattice reflections remain resolution limited before disappearing entirely above 2.84 GPa (Fig. 2a). The contrast between the multiply-split fcc Bragg peaks and the sharp superlattice reflections is proof of phase coexistence between 0.85 GPa and 2.84 GPa. We note that the onset of the high-pressure
structural phase that we observe at 0.85 GPa may explain the phase boundary at $1.3 \pm 0.4$ GPa claimed in previous neutron scattering work \cite{16}. 

\begin{figure}
\begin{center}
\includegraphics[width=3.2in]{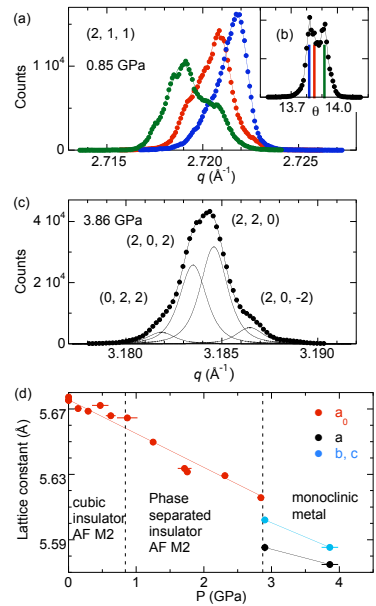}
\caption{(color online). (a) Longitudinal $\theta - 2\theta$ scans of the (2, 1, 1) lattice reflection at $0.85 \pm 0.10$ GPa, the lowest $P$ at which phase separation was observed. (b) Rocking curve recorded at $q=2.721\mathrm\AA^{-1}$ for the same (2, 1, 1) reflection. The colored bars indicate the three different central $\theta$ values used in the scans in panel a. (c) Longitudinal scan of the (2, 2, 0) lattice reflection at 3.86 GPa where only the high $P$ phase is present. The four-fold splitting of (2, 2, 0) indicates a lattice symmetry of monoclinic or lower in the metal. (d) Lattice constants vs. pressure for pure NiS$_{2}$ at $T$ = 5 K. Both the low $P$ and high $P$ phases were observed in the phase separated region, but with insufficient detail to determine the lattice parameters $a, b, c$ of the high $P$ phase. The lattice constants $a_0$ of the low $P$ phase were determined from the $d$-spacings of both the superlattice and cubic fcc reflections. }
\end{center}
\end{figure}

\section {lattice structure under pressure}
The crystal symmetry of the high-pressure metallic phase can be determined either with single crystal refinement of a single-domain specimen or with powder refinement of a polycrystalline sample with full knowledge of symmetry split peaks. We identified a single-domain sample in the high pressure phase at 2.9 GPa. Six diffraction orders were measured and a least-squares refinement reveals an almost monoclinic structure with lattice parameters (to a 95\% confidence level) $a$ = 5.5852(22) \AA, $b = c = 5.6021(6)$ \AA, $\alpha$ = 89.984(8)$^{\mathrm{o}}$, $\beta$ = 89.930(18)$^{\mathrm{o}}$, and $\gamma$ = 89.967(13)$^{\mathrm{o}}$. The values of $(a-c)/a$ and ($90^{\mathrm{o}}-\beta$) measured at $P$ = 2.9 GPa are consistent with measurements at 3.86 GPa (Fig. 3), where the four-fold splitting of the (2, 2, 0) peak only can be explained by a symmetry of monoclinic or lower. Constraining the symmetry to monoclinic, we obtain (to a 95\% confidence level) $a$ = 5.5748(9) \AA, $b = c = 5.5853(6)$ \AA,  $\beta= 89.949(1)^{\mathrm{o}}$ at 3.86 GPa. The four-fold splitting of (2, 1, 1) at 0.85 GPa (Fig. 3) is consistent with this picture, as it cannot be explained by a single phase of symmetry higher than monoclinic.

Notably, symmetry reduction on passing into the metallic phase is the opposite of what is observed in many other transition metal oxides including the prototypical Mott-Hubbard system V$_2$O$_3$, which is a rhombohedral metal and a monoclinic insulator. For a non-interacting bandstructure, a reduction of lattice symmetry favors insulating behavior \cite{5, 6}. The observation here for NiS$_2$ that the symmetry is reduced in the metal therefore emphasizes the role played by electron correlations in the insulator.

\section {Conclusion}
Our results address longstanding debates over the role of magnetism and crystal structure at the insulator-metal transition \cite{2, 3, 5, 6, 31}, while at the same time raising questions about quantum phase transitions in the presence of strong electron correlations. The broad regime of phase coexistence that we observe while tuning $U/W$ adds to a growing list of correlated electron systems that exhibit phase coexistence around a first order quantum phase transition \cite{25}. We have established that the magnetic, superexchange interaction cannot account for electron localization in the insulator and that strong electron correlations drive the insulator-metal transition even in the presence of a structural distortion. Magnetotransport measurements in the compressed metal are required to probe the evolution and gapping of the Fermi surface, as well as the role of quantum fluctuations, as the transition is approached from above. 

\acknowledgments
We are grateful to D. Robinson and J.-W. Kim for technical support at 6-ID-B of the Advanced Photon Source, and to X. Yao for growth of the crystals. The work at the University of Chicago was supported by NSF Grant No. DMR-0907025.  Use of the Advanced Photon Source was supported by the U.S. DOE-BES, under Contract No. NEAC02-06CH11357.


\begin{thebibliography}{50}        

\bibitem {2} N. F. Mott and Z. Zinamon, Rep. Prog. Phys. \textbf{33}, 881 (1970). 
\bibitem {6} M. Imada, A. Fujimori, and Y. Tokura, Rev. Mod. Phys.  \textbf{70}, 1039 (1998). 
\bibitem {24} P. W. Anderson, Phys Rev. \textbf{115}, 2 (1959).
\bibitem {3} J. C. Slater, Phys. Rev. \textbf{82}, 538 (1951); T. Moriya and K. Ueda, Rep. Prog. Phys. \textbf{66}, 1299 (2003). 
\bibitem {5} D. Adler, Rev. Mod. Phys. \textbf{40}, 714 (1968). 
\bibitem {29} Y. Feng \textit{et al.}, Rev. Sci. Instrum. \textbf{81}, 041301(2010). 
\bibitem {11} X. Yao \textit{et al.}, Phys. Rev. B \textbf{54}, 17469 (1996); J.M. Honig and J. Spa{\l}ek, Chem. Mater. \textbf{10}, 2910 (1998).

\bibitem {7} J. A. Wilson and G. D. Pitt, Philo. Mag. \textbf{23}, 1297 (1971). 
\bibitem {9} T. Miyadai \textit{et al.}, J. Phys. Soc. Jap. \textbf{38}, 115 (1975).
\bibitem {10} G. Krill \textit{et al.}, J. Phys. C \textbf{9}, 761 (1976). 
\bibitem {14} D. W. Bullett, J. Phys. C \textbf{15}, 6163 (1982).   
\bibitem {15} R. L. Kautz, M. S. Dresselhaus, D. Adler, A. Linz, Phys. Rev. B \textbf{6}, 2078 (1972). 
\bibitem {13} Q. Si, E. Abrahams, J. Dai, and J.-X. Zhu, New J. Phys. \textbf{11}, 045001 (2009). 
\bibitem {12} S. Sudo, J. Mag. Mag. Mat. \textbf{114}, 57 (1992); H. S. Jarrett \textit{et al.}, Mat. Res. Bull. \textbf{8}, 877 (1973). 
\bibitem {17} Y. Sekine \textit{et al.}, Physica B \textbf{237}, 148 (1997); N. Takeshita \textit{et al.}, arXiv:0704.0591 (2007). 
\bibitem {16} P. Panissod, G. Krill, C. Vettier, and R. Madar, Solid State Comm. \textbf{29}, 67 (1979).
\bibitem {19} N. Mori and H. Takahashi, J. Mag. Mag. Mat. \textbf{31}, 335 (1983). 
\bibitem {30} P.-G. de Gennes, Phys. Rev. \textbf{118}, 141 (1960).

\bibitem {20} T. Thio, J. W. Bennett, and T. R. Thurston, Phys. Rev. B \textbf{52}, 3555 (1995). 
\bibitem {21} H. Nagata, H. Ito, and T. Miyadai, J. Phys. Soc. Jap. \textbf{41}, 444 (1976). 
\bibitem {22} R. Werner, C. Gros, and M. Braden, Phys. Rev. B \textbf{59}, 14356 (1999).
\bibitem {32} E.R. Callen and H.B. Callen, Phys. Rev. \textbf{129}, 578 (1963). 
\bibitem {23} H. D. Lutz, J. Himmrich, B. MŸller, and G. Schneider.  J. Phys. Chem. Solids \textbf{53}, 815 (1992).

\bibitem {31} J. Spa{\l}ek, A. Datta, and J.M. Honig, Phys. Rev. Lett \textbf{59}, 728 (1987). 
\bibitem {25} Y. J. Uemura \textit{et al.}, Nature Phys. \textbf{3}, 29 (2007).


\end{thebibliography}
\end{document}